\documentclass[twoside,12pt]{article}
\usepackage{graphicx}

\newcommand{\be}{\begin{equation}}
\newcommand{\ee}{\end{equation}}
\newcommand{\bea}{\begin{eqnarray}}
\newcommand{\eea}{\end{eqnarray}}

\begin{document}

\title{ \vspace{1cm} Isospin Dynamics in Heavy Ion Collisions: from Coulomb
Barrier to Quark Gluon Plasma}
\author{M.\ Di Toro,$^{1,2}$ V.\ Baran,$^3$ M.\ Colonna,$^2$ G.\ Ferini
,$^{2}$\\ 
T.\ Gaitanos,$^4$ V.\ Giordano,$^{1,2}$ V.\ Greco,$^{1,2}$ Liu \ Bo,$^5$ 
M.\ Zielinska-Pfabe,$^6$\\ 
S.\ Plumari,$^{1,7}$ V.\ Prassa,$^8$ C.\ Rizzo,$^{1,2}$ 
J.\ Rizzo,$^{1,2}$ and H. H.\ Wolter,$^9$\\ 
\\
$^1$Dipartimento di Fisica e Astronomia dell'Universit\`a, Catania, Italy\\
$^2$INFN, Laboratori Nazionali del Sud, Catania, Italy\\
$^3$Dept. of Theoretical Physics, Bucharest Univ. and NIPNE-HH,
Romania\\
$^4$ Institute for Theoretical Physics, Giessen University, Germany\\
$^5$ IHEP, Chinese Academy of Science, Beijing, China\\
$^6$ Smith College, Northampton, Mass., USA\\
$^7$ INFN, Sezione di Catania, Italy\\
$^8$ Physics Department, Aristotles Univ.of Thessaloniki, Grece\\
$^9$ Department f\"ur Physik, Unversit\"at Munchen, Garching, Germany}
\maketitle

\begin{abstract}
Heavy Ion Collisions ($HIC$) represent a unique tool to probe the in-medium
nuclear interaction in regions away from saturation. In this report we 
present a 
selection of new reaction observables in dissipative collisions 
particularly sensitive to the
symmetry term of the nuclear Equation of State ($Iso-EoS$).
We will first discuss the Isospin Equilibration Dynamics. At low energies this
 manifests via
the recently observed Dynamical Dipole Radiation, due
to a collective neutron-proton oscillation with the symmetry term acting 
as a restoring force. At higher beam energies Iso-EoS
effects will be seen in Imbalance Ratio Measurements, in particular from the 
correlations with the total kinetic energy loss. 
For fragmentation reactions in central events we suggest to 
look at the coupling between isospin distillation and radial flow. In Neck 
Fragmentation reactions important $Iso-EoS$ information can be obtained 
from the 
correlation between isospin content and alignement.
The high density symmetry term can be probed from
 isospin effects on heavy ion reactions
at relativistic energies (few $AGeV$ range).
Rather isospin sensitive observables are proposed from nucleon/cluster 
emissions, collective flows
and meson production. The possibility  to shed light 
on the controversial neutron/proton effective mass splitting in asymmetric 
matter is also suggested. A large symmetry repulsion at high baryon density
will also lead to an ``earlier'' hadron-deconfinement transition
in n-rich matter. A suitable treatment of the isovector interaction in the
partonic $EoS$ appears very relevant.
\end{abstract}

\vskip -1.0cm
\section{Introduction: The Elusive Symmetry Term of the EoS}

The symmetry energy $E_{sym}$ appears in the energy density
$\epsilon(\rho,\rho_3) \equiv \epsilon(\rho)+\rho E_{sym} (\rho_3/\rho)^2
 + O(\rho_3/\rho)^4 +..$, expressed in terms of total ($\rho=\rho_p+\rho_n$)
 and isospin ($\rho_3=\rho_p-\rho_n$) densities. The symmetry term gets a
kinetic contribution directly from basic Pauli correlations and a potential
part from the highly controversial isospin dependence of the effective 
interactions. Both at sub-saturation and supra-saturation
densities, predictions based of the existing many-body techniques diverge 
rather widely, see \cite{fuchswci,fantoni08}. 

We  recall  that the knowledge of the 
 EoS  of asymmetric matter is very important at low densities ( e.g.  
neutron skins,
 pigmy resonances, nuclear structure at the drip lines, neutron distillation 
in fragmentation,
 neutron star formation and crust) as well as at high densities ( e.g.
 neutron star mass-radius relation, cooling, hybrid structure, transition
to a deconfined phase, formation of black holes).  
  Several observables which
are sensitive to the Iso-EoS and testable
experimentally, have been suggested
\cite{colonnaPRC57,Isospin01,baranPR,wcineck,WCI_betty,baoPR08}.
We take advantage of new opportunities in 
theory (development of rather reliable microscopic transport codes for $HIC$)
 and in experiments (availability of very asymmetric radioactive beams, 
improved possibility of measuring event-by-event correlations) to present
new results that are constraining the existing effective interaction 
models. We will discuss dissipative collisions in a wide range of beam 
energies, 
 from just above the Coulomb barrier up to the $AGeV$ range. Isospin effects
on the chiral/deconfinement transition at high baryon density will be also
discussed.
Low to Fermi energies
 will bring information on the symmetry term around (below) normal density, 
while intermediate energies will probe high density regions.
The transport codes are based on 
mean field theories, with correlations included via hard nucleon-nucleon
elastic and inelastic collisions and via stochastic forces, selfconsistently
evaluated from the mean phase-space trajectory, see 
\cite{baranPR,guarneraPLB373,colonnaNPA642,chomazPR}. 
Stochasticity is 
essential in 
order to get distributions as well as to allow the growth of dynamical 
instabilities. 

Relativistic collisions are described via a fully covariant transport 
approach, related to an effective field exchange model, where the relevant 
degrees of freedom of the nuclear dynamics are accounted for
 \cite{baranPR,liubo02,theo04,santini05,ferini05,ferini06}.
We will have a propagation of particles suitably dressed by self-energies
that will influence collective flows and in medium nucleon-nucleon inelastic 
cross sections. The construction of an $Hadron-EoS$ at high baryon and 
isospin densities will finally allow the possibility of developing a model 
of a 
hadron-deconfinement transition at high density for an asymmetric matter
\cite{ditoro_dec}. The problem of a correct treatment of the isospin in a
effective partonic $E0S$ will be stressed.  

We will always test the sensitivity of our simulation results to
different choices of the density and momentum dependence of the
Isovector part of the Equation of State ($Iso-EoS$). In the non-relativistic
frame
the potential part of the symmetry energy, $C(\rho)$, \cite{baranPR}:
\begin{equation}
\frac{E_{sym}}{A}=\frac{E_{sym}}{A}(kin)+\frac{E_{sym}}{A}(pot)\equiv 
\frac{\epsilon_F}{3} + \frac{C(\rho)}{2\rho_0}\rho
\end{equation}
 is tested by employing two different density
parametrizations, Isovector Equation of State (Iso-Eos) 
\cite{colonnaPRC57,bar02},
of the  mean field:
i) $\frac{C(\rho)}{\rho_0}=482-1638 \rho$, $(MeV fm^{3})$, for ``Asysoft'' 
EoS: ${E_{sym}/{A}}(pot)$
has a weak
density dependence close to the saturation, with an almost flat behavior below
 $\rho_0$ and even decreasing at suprasaturation; ii) a constant coefficient, 
$C=32 MeV$, for the ``Asystiff'' EoS 
choice: the interaction part of the symmetry term displays
a linear dependence with the density, i.e. with a faster decrease
at lower densities and much stiffer above saturation. The isoscalar section 
of the EoS is the same in both cases,
fixed requiring that  the saturation properties of symmetric nuclear matter
with a compressibility around $220MeV$  are reproduced.

\vskip -1.0cm
\section{Isospin Equilibration}

\subsection{The Prompt Dipole $\gamma$-Ray Emission}

The possibility of an entrance channel bremsstrahlung dipole radiation
due to an initial different N/Z distribution was suggested at the beginning
of the nineties \cite{ChomazNPA563,BortignonNPA583}. 
After several experimental evidences, in fusion as well as in deep-inelastic
reactions, \cite{PierrouPRC71,medea} and refs. therein,  
we have
now a good understanding of the process and stimulating new perspectives
from the use of radioactive beams.

During the charge equilibration process taking place
 in the first stages of dissipative reactions between colliding ions with
 different N/Z
ratios, a large amplitude dipole collective motion develops in the composite
dinuclear system, the so-called Dynamical Dipole mode. This collective dipole
gives rise to a prompt $\gamma $-ray emission which depends:
 i) on the absolute
value of the intial dipole moment
\begin{eqnarray}
&&D(t= 0)= \frac{NZ}{A} \left|{R_{Z}}(t=0)- {R_{N}}(t=0)\right| =  \nonumber \\
&&\frac{R_{P}+R_{T}}{A}Z_{P}Z_{T}\left| (\frac{N}{Z})_{T}-(\frac{N}{Z})_{P}
\right|,
\label{indip}
\end{eqnarray}
being ${R_{Z}}= \frac {\Sigma_i x_i(p)}{Z}$ and
${R_{N}}=\frac {\Sigma_i x_i(n)}{N} $ the
center of mass of protons and of neutrons respectively, while R$_{P}$ and
R$_{T}$ are the
projectile and target radii; ii) on the fusion/deep-inelastic dynamics;
 iii) on the symmetry term, below saturation, that is acting as a restoring
force.

A detailed description is obtained in mean field transport approaches,
\cite{BrinkNPA372,BaranPRL87}.
We can follow the time evolution
of the dipole moment
in the $r$-space,
 $D(t)= \frac{NZ}{A} ({R_{Z}}- {R_{N}})$ and in
$p-$space, $DK(t)=(\frac{P_{p}}{Z}-\frac{P_{n}}{N})$, 
with $P_{p}$
($P_{n}$) center of mass in momentum space for protons (neutrons),
just the canonically conjugate momentum of the $D(t)$ coordinate,
i.e. as operators $[D(t),DK(t)]=i\hbar$. 
A nice "spiral-correlation"
clearly denotes the collective nature
 of the mode, see Fig.1.

We can directly
apply a bremsstrahlung approach,
 to the dipole evolution given from the Landau-Vlasov transport
\cite{BaranPRL87}, to estimate the ``prompt'' photon emission probability
($E_{\gamma}= \hbar \omega$):
\begin{equation}
\frac{dP}{dE_{\gamma}}= \frac{2 e^2}{3\pi \hbar c^3 E_{\gamma}}
 |D''(\omega)|^{2}  \label{brems},
\end{equation}
where $D''(\omega)$ is the Fourier transform of the dipole acceleration
$D''(t)$. We remark that in this way it is possible
to evaluate, in {\it absolute} values, the corresponding pre-equilibrium
photon emission.

We must add a
couple of comments of interest for the experimental selection of the Dynamical
Dipole: i) The centroid is always shifted to lower energies (large
deformation of the dinucleus); ii) A clear angular anisotropy should be present
since the prompt mode has a definite axis of oscillation
(on the reaction plane) at variance with the statistical $GDR$.
In a  recent experiment the prompt dipole radiation has been 
investigated with
a $4 \pi$ gamma detector. A strong dipole-like photon angular distribution
$(\theta_\gamma)=W_0[1+a_2P_2(cos \theta_\gamma)]$, $\theta_\gamma$ being the 
angle between the emitted photon and  
the beam axis, has been observed,
 with the 
$a_2$ parameter close to $-1$, see \cite{medea}. 

At higher beam energies we expect a decrease of the direct dipole
radiation for two main reasons both due to the increasing importance of hard
NN collisions: i) a larger fast nucleon emission that will equilibrate the
isospin during the dipole oscillation; ii) a larger damping of the
collective mode due to $np$ collisions. 


The use of unstable neutron rich projectiles would largely increase the
effect, due to the possibility of larger entrance channel asymmetries
 \cite{ditoro_kaz07}.
In order to suggest proposals for the new $RIB$ facility $Spiral~2$, 
\cite{lewrio} we have studied fusion events in the reaction $^{132}Sn+^{58}Ni$ 
at $10AMeV$, \cite{ditoro_kaz07,spiral2}. We espect a $Monster$ 
Dynamical Dipole, the initial 
dipole moment $D(t=0)$ being of the order of 50fm, about two times the largest 
values probed so far, allowing a detailed study of the symmetry potential, 
below 
saturation,
responsible of the restoring force of the dipole oscillation and even 
of the damping,
 via the fast neutron emission.

In Figure 1 we report some global informations concerning the dipole mode
in entrance channel. 
\begin{figure}
\begin{center}
\includegraphics*[angle=-90,scale=0.33]{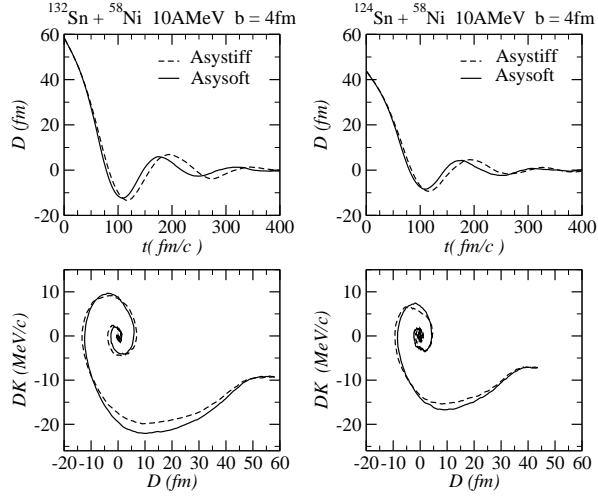}
\end{center}
\vskip -0.5cm
\caption{Dipole Dynamics at 10AMeV, $b=4fm$ centrality. 
Left Panels: Exotic ``132'' system. Upper: Time evolution of dipole moment 
D(t) in real 
space; Lower: Dipole phase-space correlation (see text).
Right Panels: same as before for the stable ``124'' system.
Solid lines correspond to Asysoft EoS, the dashed to Asystiff EoS.}
\label{dip}
\end{figure}
In the Left-Upper panel we have the time evolution of the dipole moment $D(t)$
for the ``132'' system at $b=4fm$.
We notice the large amplitude of the first oscillation
but also the delayed dynamics for the Asystiff EOS related to a weaker 
isovector
restoring force.
The phase space correlation (spiraling) between $D(t)$
and $DK(t)$, is
reported in Fig.1 (Left-Lower). 
It nicely points out a collective behavior which initiates very early,
with a dipole moment still close to the touching configuration value
reported above.
This can be explained by the fast formation of a well developed neck mean field
which sustains the collective
dipole oscillation in the dinuclear configuration.

The role of a large charge asymmetry between 
the two 
colliding nuclei can be seen from
Fig.1 (Right Panels), where we show the analogous dipole phase space 
trajectories for the stable  
$^{124}Sn+^{58}Ni$ system at the same value of impact parameter and energy. 
A clear 
reduction of the 
collective behavior is evidenced. 

\begin{figure}
\begin{center}
\includegraphics*[scale=0.33]{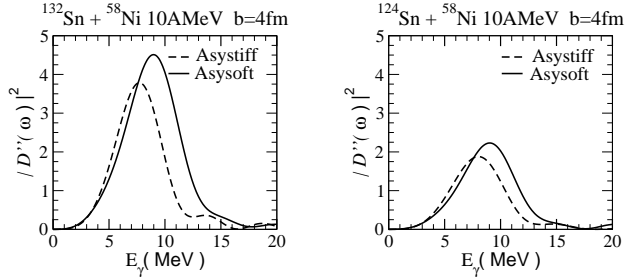}
\end{center}
\vskip -0.5cm
\caption{Left Panel, Exotic ``132'' system. Power spectra of the 
dipole acceleration at  $b=4$fm (in $c^2$ units).
Right Panel: Corresponding results for the stable ``124'' system.
Solid lines correspond to Asysoft EoS, the dashed to Asystiff EoS.}
\label{yield1}
\end{figure}

In Fig. 2(Left Panel) we report the power spectrum, 
$\mid D''(\omega) \mid^2$ in semicentral
``132'' reactions, for different $Iso-EoS$ choices.
The gamma multiplicity is simply related to it, see Eq.(\ref{brems}).
The corresponding results for the stable ``124''  system are drawn
in the Right Panel.
As expected from the larger initial charge asymmetry, we clearly see an 
increase 
of the Prompt Dipole Emission for the exotic
n-rich beam. Such entrance channel effect will be enhanced, allowing
a better observation of the Iso-EoS dependence. In fact from Fig.2 
we see also other isospin effects.

A detailed analysis can be performed just using a simple damped oscillator 
model for the dipole moment $D(t)=D(t_0) e^{i (\omega_0+i/\tau) t}$, where
$D(t_0)$ is the value at the onset of the collective dinuclear response, 
$\omega_0$ the frequency and $\tau$ the damping rate.The power spectrum of 
the dipole acceleration is given by 

\begin{equation}
\mid D''(\omega)\mid ^2 =\frac{(\omega_0^2+{1/\tau}^2)^2 {D(t_0)}^2}
 {(\omega-\omega_0)^2 + {1/\tau}^2}
\label{power}
\end{equation}

which from Eq.(\ref{brems}) leads to a total yield proportional to
$\omega_0 \tau (\omega_0^2+{1/\tau}^2){D(t_0)}^2 \simeq 
\omega_0^3 \tau {D(t_0)}^2$
since $\omega_0 \tau >1$. We remind that
in the Asystiff case we have a weaker restoring force 
for the dynamical dipole
in the dilute ``neck'' region, where the symmetry energy is smaller
\cite{baranPR}.
This is reflected in
lower values of the centroids as well as in reduced total yields, as shown 
in Fig.2. The sensitivity of $\omega_0$ to the stiffness of the symmetry
energy will be amplified by the increase of $D(t_0)$ when we use exotic,
more asymmetric beams. 

The prompt dipole radiation angular distribution is the result of the 
interplay between the collective oscillation life-time and the dinuclear 
rotation. In this sense we expect also a sensitivity to the $Iso-Eos$ of the
anusotropy, in particular for high spin event selections, \cite{dipang08}.

In the Asysoft choice we expect also larger widths of the
"resonance" due to the larger fast neutron emission.
We note the opposite effect of the Asy-stiffness on neutron vs proton 
emissions.
The latter point is important even for the possibility of an independent 
test just
measuring the $N/Z$ of the pre-equilibrium nucleon emission, \cite{pfabe_iwm}. 


\subsection{Isospin Equilibration at the Fermi Energies}

In this energy range the doorway state mechanism of the Dynamical Dipole
will disappear and so we can study a direct isospin transport in binary events.
This can be discussed in a compact way by means of the 
chemical
potentials for protons and neutrons as a function of density $\rho$ and 
isospin 
$I$ \cite{isotr05}. The $p/n$ currents can be expressed as
\begin{equation}
{\bf j}_{p/n} = D^{\rho}_{p/n}{\bf \nabla} \rho - D^{I}_{p/n}{\bf \nabla} I
\end{equation}
with $D^{\rho}_{p/n}$ the drift, and
$D^{I}_{p/n}$ the diffusion coefficients for transport, which are given 
explicitely
 in ref. \cite{isotr05}. Of interest here are the differences of currents 
between protons 
and neutrons which have a simple relation to the density dependence of the 
symmetry energy
\begin{eqnarray}
D^{\rho}_{n} - D^{\rho}_{p}  & \propto & 4 I \frac{\partial E_{sym}}
{\partial \rho} \,
 ,  \nonumber\\
D^{I}_{n} - D^{I}_{p} & \propto & 4 \rho E_{sym} \, .
\label{trcoeff}
\end{eqnarray}
Thus the isospin transport due to density gradients, i.e. isospin migration, 
depends on the slope of the symmetry energy, or the symmetry pressure, 
while the 
transport due to isospin concentration gradients, i.e. isospin diffusion, 
depends on
 the absolute value of the symmetry energy. 

 We can discuss the asymmetries of the various parts of the reaction system
(gas,  PLF/TLF's , and in  the  case of ternary events,  IMF's ).
 In particular, we study  the
 so-called Imbalance Ratio 
\cite{imbalance}, which is defined as
\begin{equation}
R^x_{P,T} = \frac{2(x^M-x^{eq})}{(x^H-x^L)}~,
\label{imb_rat}
\end{equation}
with $x^{eq}=\frac{1}{2}(x^H+x^L)$.
 Here, $x$ is an isospin sensitive quantity
that has to be investigated with respect to
equilibration.   In this work we consider primarily the asymmetry 
$\beta=(N-Z)/(N+Z)$,
but also other quantities, such as isoscaling coefficients, ratios of 
production of light
 fragments, etc, can be of interest \cite{WCI_betty}. 
The indices $H$ and $L$ refer to the symmetric reaction
between the
heavy  ($n$-rich) and the light ($n$-poor)  systems, while $M$ refers to the
mixed reaction.
$P,T$ denote the rapidity region, in which this quantity is measured, in
particular the
PLF and TLF rapidity regions. Clearly, this ratio is $\pm1$ in
the projectile
and target regions, respectively, for complete transparency, and oppositely
for complete
rebound, while it is zero for complete equilibration.

\begin{figure}[t]
\centering
\includegraphics[width=7.0cm]{erice3a.eps}
\includegraphics[width=7.5cm]{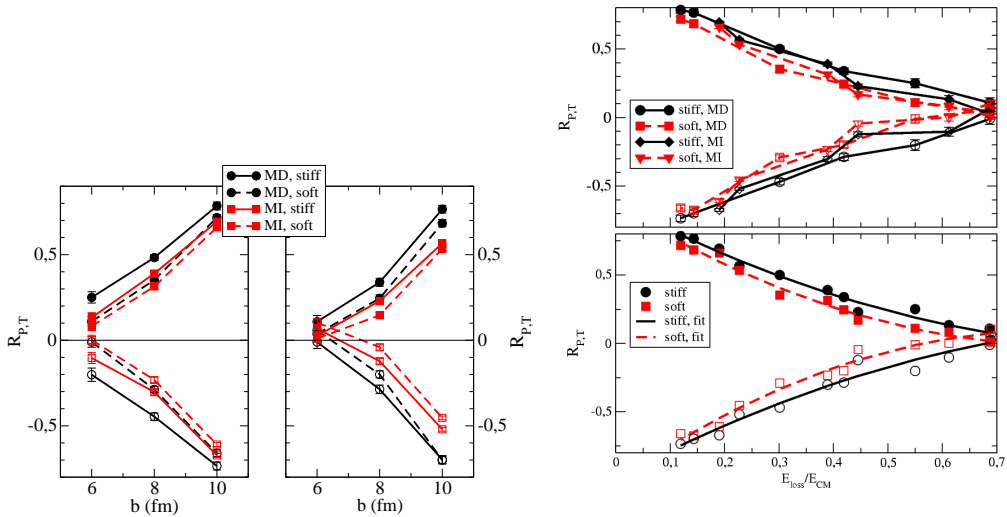}
\caption{ Left Panel.Imbalance ratios for $Sn + Sn$ collisions for 
incident energies
of 50 (left) 
and 35 $AMeV$ (right) as a function of the impact parameter. Signatures of 
the curves: 
iso-EoS stiff (solid lines), soft (dashed lines); MD interaction (circles),
 MI interaction (squares); projectile rapidity ( full symbols, upper curves ),
 target rapidity ( open symbols, lower curves ).
Right Panel. Imbalance ratios as a function of relative energy loss for both 
beam energies. 
Upper: Separately for 
stiff (solid) and soft (dashed) iso-EoS, and for MD 
(circles and squares) and MI 
(diamonds and triangles) interactions, in the projectile region (full symbols)
 and the target region 
(open symbols).
Lower: Quadratic fit to all points for the stiff (solid), resp.
 soft (dashed) 
iso-EoS.}
\label{imb_eloss}
\end{figure}

In a simple model we can show that the imbalance ratio mainly depends on two
quantities: the strength of the symmetry energy and the interaction
time between the two reaction partners.
Let us take, for instance, the asymmetry $\beta$ of the PLF (or TLF) as the
quantity $x$.
As a first order approximation, in the mixed reaction this quantity relaxes
towards
its complete equilibration value, $\beta_{eq} = (\beta_H + \beta_L)/2$ as
\begin{equation}
\label{dif_new}
\beta^M_{P,T} = \beta^{eq} + (\beta^{H,L} -  \beta^{eq})~e^{-t/\tau},
\end{equation}
where $t$ is the time elapsed while the reaction partners are interacting
(interaction time) and the damping $\tau$ is mainly connected to the strength 
of the symmetry energy. 
Inserting this expression into Eq.(\ref{imb_rat}), one obtains
$ R^{\beta}_{P,T} = \pm e^{-t/\tau}$ for the PLF and TLF regions, respectively.
Hence the imbalance ratio can be considered as a good observable to 
trace back the strength
of the symmetry energy from the reaction dynamics
provided a suitable selection of the interaction time is performed.
 


The centrality dependence of the Imbalance Ratio, for (Sn,Sn) collisions,
has been investigated in experiments \cite{tsang92} as well as in theory
\cite{isotr05,BALi}. We propose here a new analysis
 which appears experimentally more selective. 
The interaction time certainly
 influences the amount of  isospin equilibration, see Eq.(6) and
refs. \cite{isotr05,isotr07}.
Longer interaction times should be correlated to
a larger 
dissipation. It is then natural to look at the correlation between
the imbalance ratio and the total kinetic energy loss.
In this way we can also better disentangle dynamical effects of the isoscalar 
and isovector part of the EoS, see \cite{isotr07}.

It is seen in Fig.3 that the curves for the 
{\it asy-soft} EoS (dashed) are 
generally lower in the projectile region
 (and oppositely for the target region), i.e. show 
more equilibration, than those for the {\it asy-stiff} EoS. In order 
to emphasize 
this trend we have, in 
the lower panel of the figure, collected together  all  the values for the
stiff (circles) and 
the soft (squares) iso-EoS, and fitted them by a quadratic curve. 
It is seen that this fit 
gives a good representation of the trend of the results.

The difference between the curves for the stiff and soft iso-EOS in the 
lower panel then isolates
 the influence of the iso-EoS from kinematical effects 
associated with the 
interaction time. It is seen,
 that there is a systematic effect of the symmetry energy of the order 
of about 20 percent, 
which should be measurable. The correlation suggested in Fig.3
should represent 
a general feature of isospin diffusion, and it would be of great 
interest to verify 
experimentally.

\begin{figure}[t]
\centering
\includegraphics[width=7.0cm]{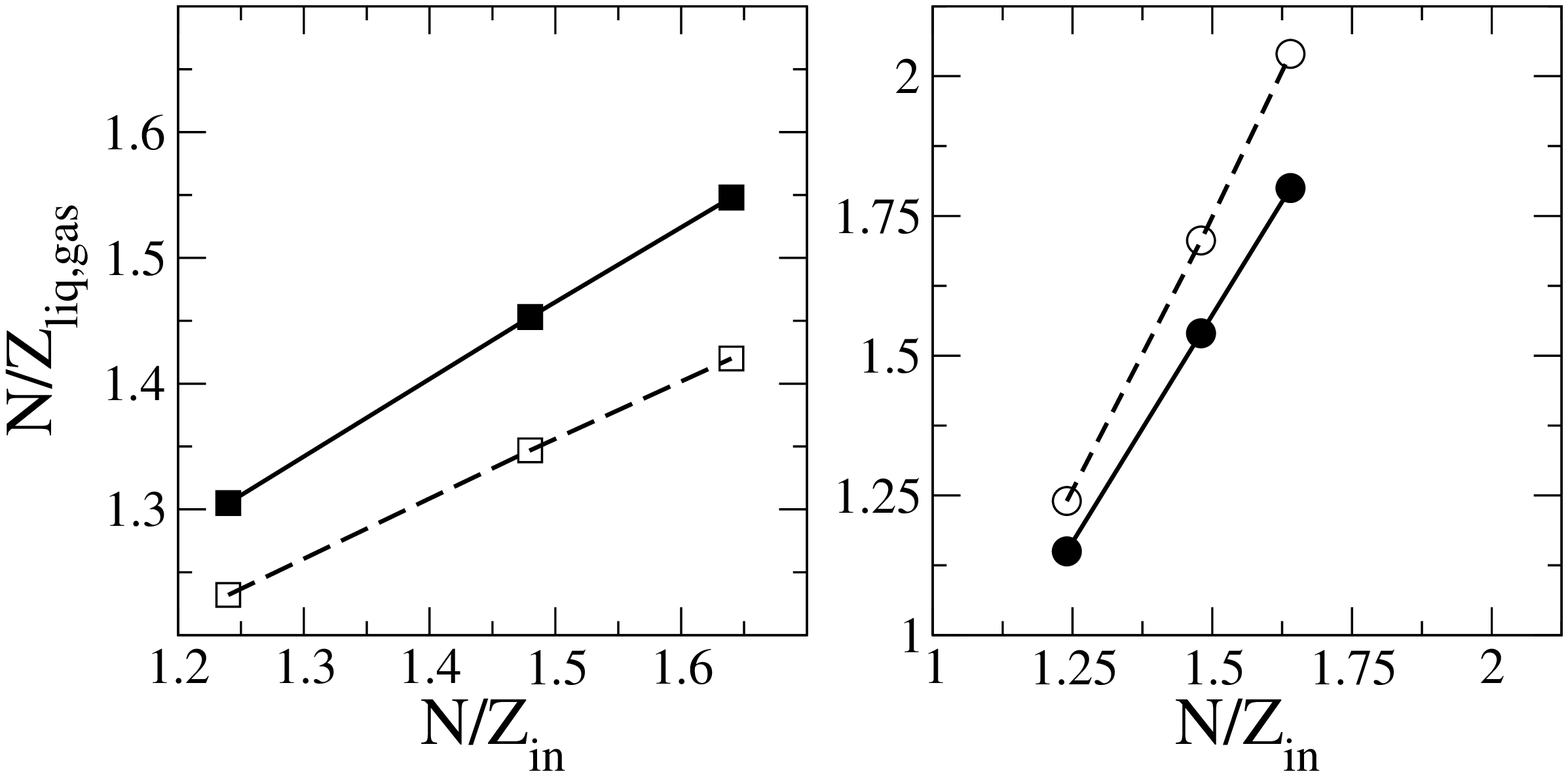}
\hskip 0.5cm
\includegraphics[width=7.0cm]{erice4b.eps}
\caption{Left Panel.The N/Z of the liquid (left) and of the gas (right) phase
is displayed as a function of the system initial N/Z.
Full lines and symbols refer to the asystiff parameterization. Dashed
lines and open symbols are for asysoft.
Right Panel.
The fragment N/Z (see text) as a function of the kinetic energy.
Left: Asystiff;  Right: Asysoft.
}
\label{iso_kin}
\end{figure}

\vskip -1.0cm
\section{Isospin Distillation with Radial Flow}

In central collisions at 30-50 MeV/A, where the full disassembly of the system
into many fragments is observed, one can study specifically properties of
liquid-gas phase transitions occurring in asymmetric matter 
\cite{mue95,bao197,BaranNPA632,chomazPR,baranPR}. 
For instance,
in neutron-rich matter, phase co-existence leads to
a different asymmetry
in the liquid and gaseous phase:  fragments (liquid) appear more symmetric
with respect to the initial matter, while light particles (gas) are
more neutron-rich.
The amplitude of this effect
depends on
 specific properties of the isovector part of the nuclear interaction,
namely on the value and the derivative of the symmetry
energy at low density.  

This investigation is interesting in a more general context:
In heavy ion collisions the dilute phase appears during the expansion
of the interacting matter.
Thus we study effects of the coupling of expansion, fragmentation and 
distillation in a 
two-component (neutron-proton) system \cite{col07} .

We focus on central collisions, $b = 2~fm$, considering symmetric 
reactions
between systems having three different initial asymmetry:
$^{112}Sn + ^{112}Sn,^{124}Sn + ^{124}Sn,
^{132}Sn + ^{132}Sn,$ with $(N/Z)_{in}$ = 1.24,1.48,1.64, respectively.
The considered
beam energy is 50 MeV/A.
1200 events have been run for each reaction and for each of the two
symmetry energies adopted (asy-soft and asystiff, see before)
\cite{col07}.
The average N/Z of emitted nucleons (gas phase) and Intermediate Mass 
Fragments ($IMF$)
is presented in Fig.4 (Left Panel) as a function of the initial $(N/Z)_{in}$
of the three colliding systems.
One observes a clear Isospin-Distillation effect, i.e.
the gas phase (right) more neutron-rich than the IMF's (left).
This is particular evident in the
Asysoft case
due to the larger value of the symmetry energy at low density
\cite{baranPR}.

In central collisions, after the initial collisional shock, the system
expands and breaks up into many pieces, due to the development of volume
(spinodal) and surface instabilities.  The formation of
a bubble-like configuration is observed, where the initial
fragments are located \cite{colonna_npa742}.

Fragmentation originates from the break-up of a composite source
that expands with a given velocity field.
Since neutrons and protons experience different forces,
one may expect a different radial flow for the two species.
In this case,  the N/Z composition of the source would not be uniform,
but would depend on the radial distance from the center or mass or,
equivalently, on the local velocity.
This trend should then be reflected in a clear correlation between
isospin content and kinetic energy of the formed $IMF$'s, \cite{col07}.

This observable is plotted in Fig.4 (Right Panel) for the three reactions.
The behaviour observed is rather sensitive to
the Iso-EoS.
For the proton-rich system, the N/Z decreases with the
fragment kinetic energy, expecially in the Asystiff case (left), where the
symmetry energy is relatively small at low density.
In this case, the Coulomb repulsion
pushes the protons towards the surface of the system. Hence, more
symmetric fragments acquire larger velocity.
The decreasing trend is less pronounced in the Asysoft case
(right) because Coulomb effects on protons
are counterbalanced by the larger
attraction of the symmetry potential.
In systems with higher initial asymmetry, the decreasing
trend is inversed, due to the larger neutron repulsion in neutron-rich
systems.

In conclusion, this analysis reveals the existence of
significant, EoS-dependent correlations between
the $N/Z$ and the kinetic energy of
IMF's produced in central collisions.

\begin{figure}
\vskip -0.5cm
 \includegraphics[scale=0.28]{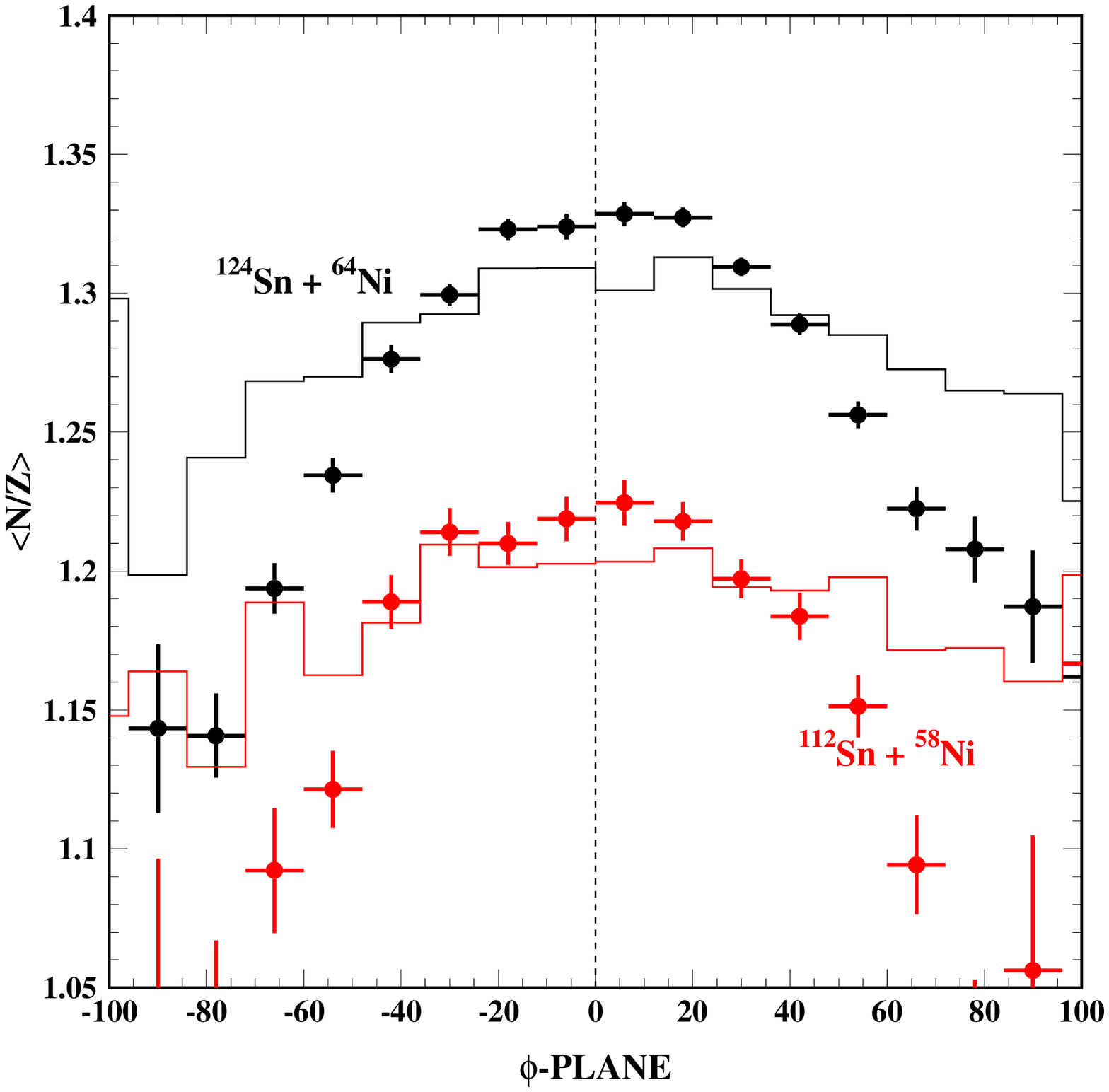}
\begin{picture}(0,0)
\put(6.,140.){\mbox{\includegraphics[scale=0.30,angle=-90]{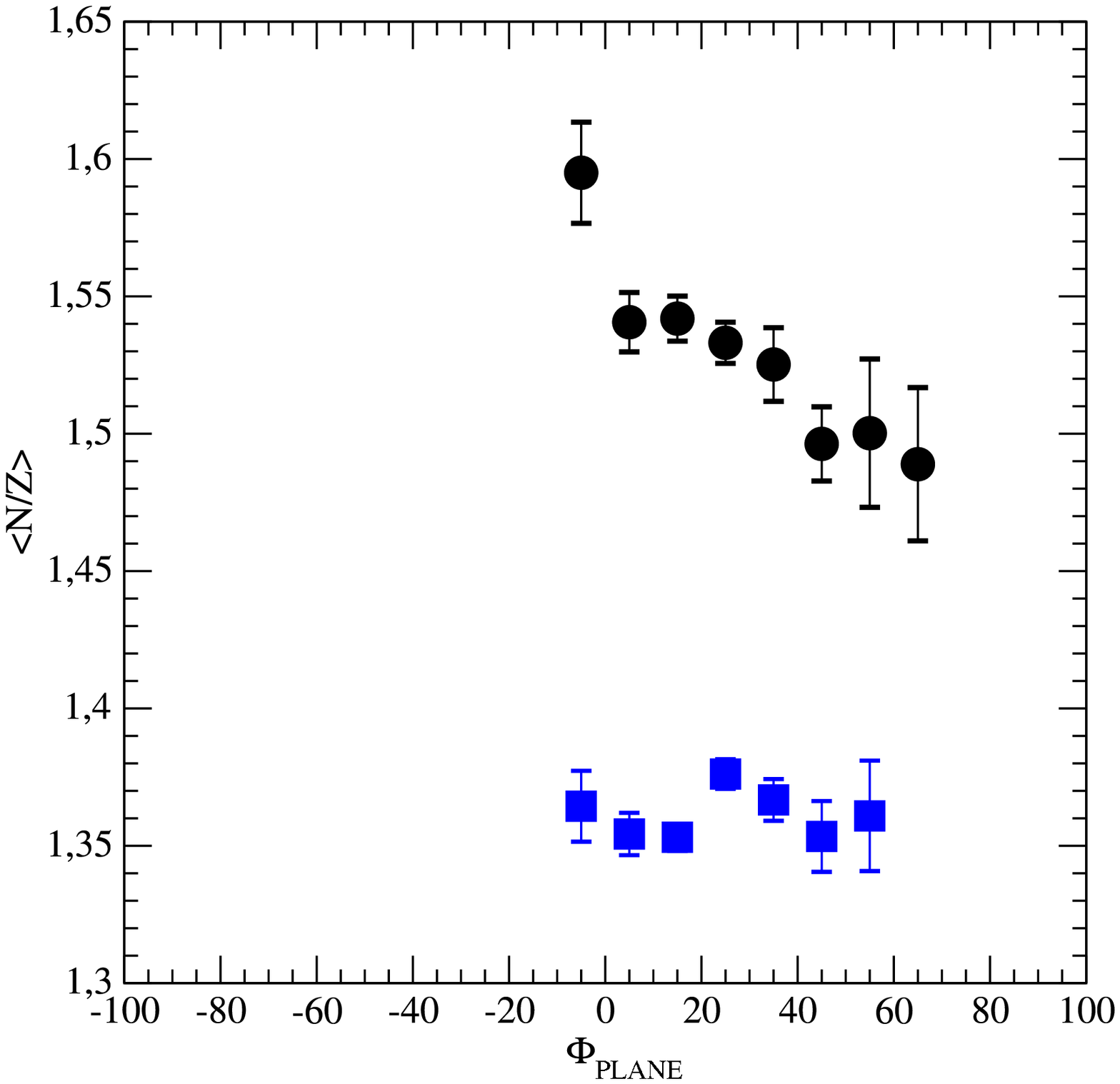}}}
\end{picture}
\caption{Correlation between $N/Z$ of $IMF$ and $alignement$ in ternary
events of  the $^{124}Sn+^{64}Ni$
reaction at $35~AMeV$. $Left~Panel$. Exp. results: points correspond to fast 
formed $IMF$s (Viola-violation selection); histogram for all $IMF$s at 
mid-rapidity (including statistical emissions). $Right~Panel$. Simulation 
results: squares, asysoft; circles, asystiff.}  
\label{nzphi}
\end{figure}

\vskip -1.0cm
\section{Isospin Dynamics in Neck Fragmentation at Fermi Energies}

It is now quite well established that the largest part of the reaction
cross section for dissipative collisions at Fermi energies goes
through the {\it Neck Fragmentation} channel, with $IMF$s directly
produced in the interacting zone in semiperipheral collisions on very short
time scales \cite{colonnaNPA589,wcineck}. We can predict interesting 
isospin transport 
effects for this new
fragmentation mechanism since clusters are formed still in a dilute
asymmetric matter but always in contact with the regions of the
projectile-like and target-like remnants almost at normal densities.
As discussed in Sect.2.2 in presence of density gradients the isospin transport
is mainly ruled by drift coefficients and so
we expect a larger neutron flow to
 the neck clusters for a stiffer symmetry energy around saturation, 
\cite{baranPR,baranPRC72}. The isospin dynamics can be directly extracted 
from correlations between $N/Z$, $alignement$ and emission times of the $IMF$s.
The alignment between $PLF-IMF$ and $PLF-TLF$ directions
represents a very convincing evidence of the dynamical origin of the 
mid-rapidity fragments produced on short time scales \cite{baranNPA730}. 
The form of the
$\Phi_{plane}$ distributions (centroid and width) can give a direct
information on the fragmentation mechanism \cite{dynfiss05}. Recent 
calculations confirm that the light fragments are emitted first, a general 
feature expected for that rupture mechanism \cite{liontiPLB625}. 
The same conclusion can be derived from {\it direct} emission time 
measurements based on deviations from Viola systematics  observed
in event-by-event velocity correlations between $IMF$s and the $PLF/TLF$ 
residues
 \cite{baranNPA730,dynfiss05,velcorr04}. 
 We can figure out
   a continuous transition from fast produced fragments via neck instabilities
   to clusters formed in a dynamical fission of the projectile(target) 
   residues up to the evaporated ones (statistical fission). Along this 
   line it would be even possible to disentangle the effects of volume
   and shape instabilities. 
A neutron enrichment of the overlap ("neck") region is
   expected, due to the neutron migration from higher (spectator) to 
   lower (neck) density regions, directly related to
   the slope of the symmetry energy \cite{liontiPLB625}.

A very nice new analysis has been performed on the $Sn+Ni$ data at $35~AMeV$
by the Chimera Collab.,\cite{defilposter},
 see Fig.\ref{nzphi} left panel.
A strong correlation between neutron enrichement and alignement (when the 
short emission time selection is enforced) is seen, that can be reproduced 
only with 
a stiff behavior of the symmetry energy, Fig.3 right panel 
(for primary fragments) \cite{baran08}. 
This represents a 
clear evidence in favor of a relatively large slope (symmetry pressure) 
around saturation. We note a recent confirmation from structure data,
i.e. from monopole resonances in Sn-isotopes \cite{garg_prl07}.

\vskip -1.0cm
\section{Isospin Effects at High Baryon Density: Effective Mass Splitting and 
Collective Flows}

\begin{figure}
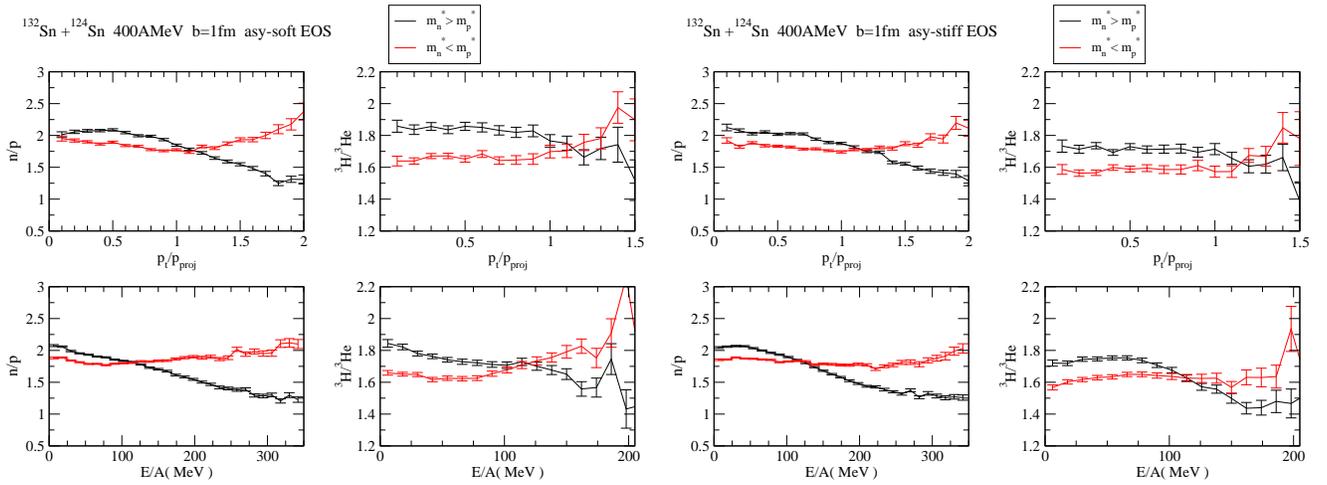

\centering
\includegraphics[width=8.5cm]{erice6a.eps}
\hskip 0.2cm
\includegraphics[width=8.5cm]{erice6b.eps}
\caption{132Sn+124Sn at 400AMeV, central coll.Isospin content of 
nucleon and light ion emissions vs $p_t$ (upper) and kinetic energy (lower).
Upper Panel:
Asysoft; Lower Panel: Asystiff.}
\label{fastratios}
\end{figure}

The problem of Momentum Dependence in the Isovector
channel ($Iso-MD$) is still very controversial and it would be extremely
important to get more definite experimental information,
see the recent refs. 
\cite{BaoNPA735,ditoroAIP05,rizzoPRC72}. 
Exotic Beams at intermediate energies are
of interest in order to have high momentum particles and to test regions
of high baryon (isoscalar) and isospin (isovector) density during the
reaction dynamics.
Our transport code has been implemented with 
a $BGBD-like$ \cite{GalePRC41,BombaciNPA583}   mean field 
with a different $(n,p)$ momentum dependence, see 
\cite{ditoroAIP05,rizzoPRC72}. This will allow to follow the dynamical
effect of opposite n/p effective mass splitting while keeping the
same density dependence of the symmetry energy \cite{isotr07}.

We present here some preliminary results for reactions induced by $^{132}Sn$
 beams on $^{124}Sn$ targets at $400AMeV$ \cite{vale08}. For central 
collisions in the interacting zone we can reach baryon densities about
$1.7-1.8 \rho_0$ in a transient time of the order of 15-20 fm/c. The system 
is quickly expanding and the Freeze-Out time is around 50fm/c.

In Fig.6 we present the  $(n/p)$ and $^3H/^3He$ yield ratios at freeze-out,
for two choices of Asy-stiffness and mass splitting, vs. transverse
momentum (upper curves) and kinetic energy (lower curves). In this way we can
separate particle emissions from sources at different densities.
We note two interesting features: i) the curves are crossing at
$p_t \simeq p_{projectile}= 2.13 fm^{-1}$; ii) the effect is not much dependent
on the stiffness of the symmetry term. The crossing nicely corresponds
to a source at baryon density $\rho \simeq 1.6 \rho_0$, 
 \cite{ditoroAIP05,rizzoPRC72}. These data seem to be suitable
to disentangle $Iso-MD$ effects.

Collective flows are very good candidates since they are 
expected to be 
very sensitive to the momentum
dependence of the mean field, see \cite{DanielNPA673,baranPR}.
The transverse flow, 
$V_1(y,p_t)=\langle \frac{p_x}{p_t} \rangle$,
provides information on the anisotropy of 
nucleon emission on the reaction plane.
Very important for the reaction dynamics is the elliptic
flow,
$V_2(y,p_t)=\langle \frac{p_x^2-p_y^2}{p_t^2} \rangle$.
 The sign of $V_2$ indicates the azimuthal anisotropy of emission:
on the reaction
plane ($V_2>0$) or out-of-plane ($squeeze-out,~V_2<0$)
\cite{DanielNPA673}.
We have then tested the $Iso-MD$ of the fields
just evaluating the $Difference$ of neutron/proton transverse and elliptic 
flows 
$
V^{(n-p)}_{1,2} (y,p_t) \equiv V^n_{1,2}(y,p_t) - V^p_{1,2}(y,p_t)
$ 
at various rapidities and transverse momenta in semicentral
($b/b_{max}=0.5$) $^{197}Au+^{197}Au$ collisons at $400AMeV$.
For the nucleon elliptic flows
 the mass splitting
effect is evident at all rapidities, and nicely increasing at larger
rapidities and transverse momenta, with more neutron flow when
$m_n^*<m_p^*$.
\begin{figure}[t]
\centering
\includegraphics[width=7.0cm]{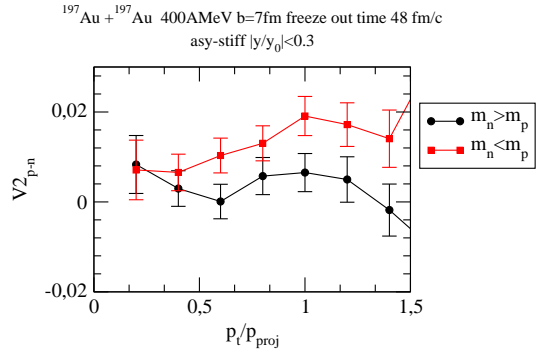}
\caption{
Transverse momentum dependence of the difference between proton 
and neutron $V_2$ flows, at mid-rapidity, in a 
semi-central
reaction Au+Au at 400AMeV.} 
\label{v2dif}
\vskip -0.3cm
\end{figure}
From Fig.7 we clearly see how at mid-rapidity the mass splitting effects 
are more evident
for higher tranverse momentum selections, i.e. for high
density sources. In particular
the elliptic flow difference becomes negative when
$m_n^*<m_p^*$, revealing a faster neutron emission and so more neutron
squeeze out (more spectator shadowing).
In correspondance the proton flow
is more negative (more proton squeeeze out) when $m_n^*>m_p^*$.
It is however difficult to draw definite conclusions only from proton data.
The measurement of n/p flow differences appears essential.
Due to the difficulties in
measuring neutrons, our suggestion is to measure the difference between
light isobar flows, like $^3H$ vs. $^3He$ and so on.
We expect to still see effective mass splitting effects. 

\section{Isospin Effects on Meson Production in Relativistic Heavy Ion
Collisions}
The phenomenology of isospin effects on heavy ion reactions
at intermediate energies (few $AGeV$ range) is extremely rich and can allow
a ``direct'' study of the covariant structure of the isovector interaction
in a high density hadron medium. We work within a relativistic transport frame,
beyond the cascade picture, 
 consistently derived from effective Lagrangians, where isospin effects
are accounted for in the mean field and collision terms.
We show that rather sensitive observables are provided by the 
pion/kaon production ($\pi^-/\pi^+$, 
$K^0/K^+$ yields). Relevant non-equilibrium effects
are stressed.

An effective Lagrangian approach to the hadron interacting system is
extended to the isospin degree of freedom: within the same frame equilibrium
properties ($EoS$, \cite{qhd}) and transport dynamics 
can be consistently derived.
Within a covariant picture of the nuclear mean field, 
 for the description of the symmetry energy at saturation
($a_{4}$ parameter of the Weizs\"{a}ecker mass formula)
(a) only the Lorentz vector $\rho$ mesonic field, 
and (b) both, the vector $\rho$ (repulsive) and  scalar 
$\delta$ (attractive) effective 
fields are included. 
In the latter case a rather intuitive form of the Symmetry Energy can be
obtained \cite{liubo02,theo04}
\begin{equation} 
E_{sym} = \frac{1}{6} \frac{k_{F}^{2}}{E_{F}} +  
\frac{1}{2} 
\left[ f_{\rho} - f_{\delta}\left( \frac{m^{*}}{E_{F}} \right)^{2} 
\right] \rho_{B}. 
\label{esym3} 
\quad . 
\end{equation} 
The competition between scalar and vector fields leads
to a stiffer symmetry term at high density \cite{liubo02,baranPR}. 

We present
here observable effects in the dynamics of heavy ion 
collisions. 
We focus our attention on the isospin content of meson 
production.
The starting point is
a simple phenomenological version of the Non-Linear (with respect to the 
iso-scalar, Lorentz scalar $\sigma$ field) effective nucleon-boson 
field theory,  
the Quantum-Hadro-Dynamics \cite{qhd}. 
According to this picture 
the presence of the hadronic medium leads to effective masses and 
momenta $M^{*}=M+\Sigma_{s}$,   
 $k^{*\mu}=k^{\mu}-\Sigma^{\mu}$, with
$\Sigma_{s},~\Sigma^{\mu}$
 scalar and vector self-energies. 
For asymmetric matter the self-energies are different for protons and 
neutrons, depending on the isovector meson contributions. 
We will call the 
corresponding models as $NL\rho$ and $NL\rho\delta$, respectively, and
just $NL$ the case without isovector interactions. 
For the more general $NL\rho\delta$ case  
the self-energies 
of protons and neutrons read:
\begin{eqnarray}
\Sigma_{s}(p,n) = - f_{\sigma}\sigma(\rho_{s}) \pm f_{\delta}\rho_{s3}, 
\nonumber \\
\Sigma^{\mu}(p,n) = f_{\omega}j^{\mu} \mp f_{\rho}j^{\mu}_{3},
\label{selfen}
\end{eqnarray}
(upper signs for neutrons), where $\rho_{s}=\rho_{sp}+\rho_{sn},~
j^{\alpha}=j^{\alpha}_{p}+j^{\alpha}_{n},\rho_{s3}=\rho_{sp}-\rho_{sn},
~j^{\alpha}_{3}=j^{\alpha}_{p}-j^{\alpha}_{n}$ are the total and 
isospin scalar 
densities and currents and $f_{\sigma,\omega,\rho,\delta}$  are the coupling 
constants of the various 
mesonic fields. 
$\sigma(\rho_{s})$ is the solution of the non linear 
equation for the $\sigma$ field \cite{liubo02,baranPR}.
From the form of the scalar self-energies we note that in n-rich environment
the neutron effective masses are definitely below the proton ones.

For the description of heavy ion collisions we solve
the covariant transport equation of the Boltzmann type 
within the 
Relativistic Landau
Vlasov ($RLV$) method, using phase-space Gaussian test particles 
\cite{FuchsNPA589},
and applying
a Monte-Carlo procedure for the hard hadron collisions.
The collision term includes elastic and inelastic processes involving
the production/absorption of the $\Delta(1232 MeV)$ and $N^{*}(1440
MeV)$ resonances as well as their decays into pion channels,
 \cite{ferini05}.

Kaon production has been proven to be a reliable observable for the
high density $EoS$ in the isoscalar sector 
\cite{FuchsPPNP56,HartPRL96}.
Here we show that the $K^{0,+}$
production (in particular the $K^0/K^+$ yield ratio) can be also used to 
probe the isovector part of the $EoS$,
\cite{ferini06,Pra07}.

Using our $RMF$ transport approach  we analyze 
pion and kaon production in central $^{197}Au+^{197}Au$ collisions in 
the $0.8-1.8~AGeV$
 beam 
energy range, comparing models giving the same ``soft'' $EoS$ for symmetric 
matter and with different effective field choices for 
$E_{sym}$. 

\begin{figure}[t] 
\begin{center}
\includegraphics[scale=0.32]{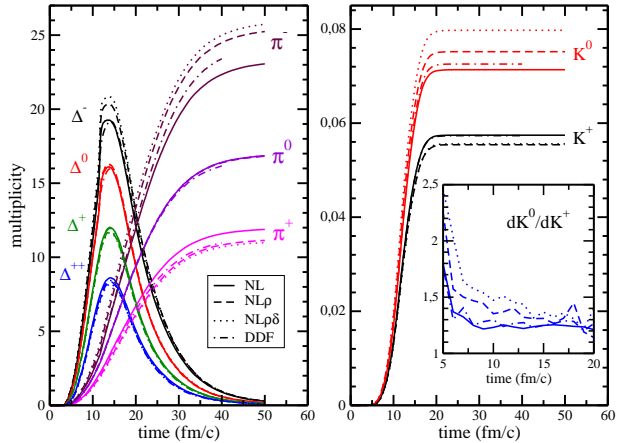} 
\vskip -0.3cm
\caption{Time evolution of the $\Delta^{\pm,0,++}$ resonances
and pions $\pi^{\pm,0}$ 
(left),  and  kaons ($K^{+,0}$
 (right) for a central ($b=0$ fm impact parameter)  
Au+Au collision at 1 AGeV incident energy. Transport calculation using the  
$NL, NL\rho, NL\rho\delta$ and $DDF$ models for the iso-vector part of the  
nuclear $EoS$ are shown. The inset contains the differential $K^0/K^+$  ratio
as a function of the kaon emission time.  
}
\vskip -1.0cm
\label{kaon1} 
\end{center}
\end{figure}

Fig. \ref{kaon1} reports  the temporal evolution of $\Delta^{\pm,0,++}$  
resonances, pions ($\pi^{\pm,0}$) and kaons ($K^{+,0}$)  
for central Au+Au collisions at $1AGeV$.

It is clear that, while the pion yield freezes out at times of the order of 
$50 fm/c$, i.e. at the final stage of the reaction (and at low densities),
kaon production occurs within the very early (compression) stage,
 and the yield saturates at around $20 fm/c$. 
From Fig. \ref{kaon1} we see that the pion results are  
weakly dependent on the  
isospin part of the nuclear mean field.
However, a slight increase (decrease) in the $\pi^{-}$ ($\pi^{+}$) 
multiplicity is observed when going from the $NL$ to the 
$NL\rho$ and then to
the $NL\rho\delta$ model, i.e. increasing the vector contribution $f_\rho$
in the isovector channel. This trend is 
more pronounced for kaons, see the
right panel, due to the high density selection of the source and the
proximity to the production threshold. Consistently, as shown in the
insert, larger effects are expected for early emitted kaons, reflecting the 
early $N/Z$ of the system. 

When isovector fields are included the symmetry potential energy in 
neutron-rich matter is repulsive for neutrons and attractive for protons.
In a $HIC$ this leads to a fast, pre-equilibrium, emission of neutrons.
 Such a $mean~field$ mechanism, often referred to as isospin fractionation
\cite{baranPR}, is responsible for a reduction of the neutron
to proton ratio during the high density phase, with direct consequences
on particle production in inelastic $NN$ collisions.
$Threshold$ effects represent a more subtle point. The energy 
conservation in
a hadron collision in general has to be formulated in terms of the canonical
momenta, i.e. for a reaction $1+2 \rightarrow 3+4$ as
$
s_{in} = (k_1^\mu + k_2^\mu)^2 = (k_3^\mu + k_4^\mu)^2 = s_{out}.
$
Since hadrons are propagating with effective (kinetic) momenta and masses,
 an equivalent relation should be formulated starting from the effective
in-medium quantities $k^{*\mu}=k^\mu-\Sigma^\mu$ and $m^*=m+\Sigma_s$, where
$\Sigma_s$ and $\Sigma^\mu$ are the scalar and vector self-energies,
Eqs.(\ref{selfen}).
The self-energy contributions will influence the particle production at the
level of thresholds as well as of the phase space available in the final 
channel. In fact the {\it threshold} effect is dominant and consequently the
results are nicely sensitive to the covariant structure of the isovector
fields.
At each beam energy we see an
increase of the $\pi^-/\pi^+$ and 
$K^{0}/K^{+}$ 
yield ratios with the models
$NL \rightarrow DDF \rightarrow NL\rho \rightarrow NL\rho\delta$. 
The effect is larger for the $K^{0}/K^{+}$ compared to the $\pi^-/\pi^+$
ratio. This is due to the subthreshold production and to the fact that
the isospin effect enters twice in the two-step production of kaons, see
\cite{ferini06}. 
Interestingly the Iso-$EoS$ effect for pions is increasing at lower energies,
when approaching the production threshold.

We have to note that in a previous study of kaon production in excited nuclear
matter the dependence of the $K^{0}/K^{+}$ yield ratio on the effective
isovector interaction appears much larger (see Fig.8 of 
ref.\cite{ferini05}).
The point is that in the non-equilibrium case of a heavy ion collision
the asymmetry of the source where kaons are produced is in fact reduced
by the $n \rightarrow p$ ``transformation'', due to the favored 
$nn \rightarrow p\Delta^-$ processes. This effect is almost absent at 
equilibrium due to the inverse transitions, see Fig.3 of 
ref.\cite{ferini05}. Moreover in infinite nuclear matter even the fast
neutron emission is not present. 
This result clearly shows that chemical equilibrium models can lead to
uncorrect results when used for transient states of an $open$ system.

\section{On the Transition to a Mixed Hadron-Quark Phase at High
Baryon and Isospin Density}
The possibility of the transition to a mixed hadron-quark phase, 
at high baryon and isospin density, is finally suggested. Some signatures
could come from an expected ``neutron trapping'' effect.
\begin{figure}
\begin{center}
\includegraphics[scale=0.34]{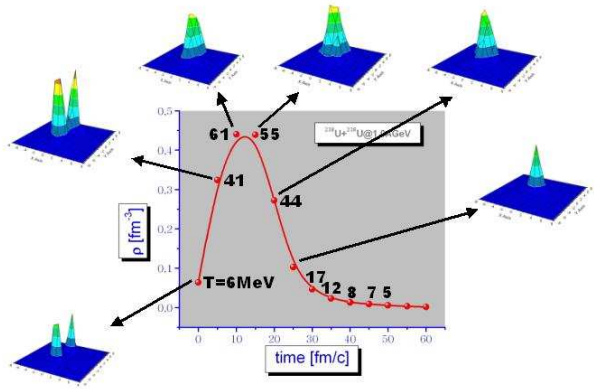}
\caption{$^{238}U+^{238}U$, $1~AGeV$, semicentral. Correlation between 
density, 
temperature (black values), momentum
thermalization (3-D plots), inside a cubic cell, 2.5 $fm$ wide, in the center
of mass of the system.}
\label{figUU}
\end{center}
\end{figure}

In order to check the possibility of observing some precursor signals
of a new physics even in collisions of stable nuclei at
intermediate energies we have performed some event simulations for the
collision of very heavy, neutron-rich, elements. We have chosen the
reaction $^{238}U+^{238}U$ (average proton fraction $Z/A=0.39$) at
$1~AGeV$ and semicentral impact parameter $b=7~fm$ just to increase
the neutron excess in the interacting region. 
In  Fig.~\ref{figUU} we report the evolution of momentum distribution
and baryon density in a space cell located in the c.m. of the system.
We see that after about $10~fm/c$ a local
equilibration is achieved.  We have a unique Fermi distribution and
from a simple fit we can evaluate the local temperature 
(black numbers in MeV).
We note that a rather exotic nuclear matter is formed in a transient
time of the order of $10~fm/c$, with baryon density around $3-4\rho_0$,
temperature $50-60~MeV$, energy density $500~MeV~fm^{-3}$ and proton
fraction between $0.35$ and $0.40$, likely inside the estimated mixed 
phase region.

In fact we can study the isospin dependence of the transition densities
\cite{ditoro_dec}.
The structure of the mixed phase is obtained by
imposing the Gibbs conditions \cite{Landaustat} for
chemical potentials and pressure and by requiring
the conservation of the total baryon and isospin densities
\begin{eqnarray}\label{gibbs}
&&\mu_B^{(H)} = \mu_B^{(Q)}\, ,~~  
\mu_3^{(H)} = \mu_3^{(Q)} \, ,  \nonumber \\
&&P^{(H)}(T,\mu_{B,3}^{(H)}) = P^{(Q)} (T,\mu_{B,3}^{(Q)})\, ,\nonumber \\
&&\rho_B=(1-\chi)\rho_B^H+\chi\rho_B^Q \, , \nonumber \\
&&\rho_3=(1-\chi)\rho_3^H+\chi\rho_3^Q\, , 
\end{eqnarray}
where $\chi$ is the fraction of quark matter in the mixed phase.
\begin{figure}
\begin{center}
\includegraphics[angle=+90,scale=0.30]{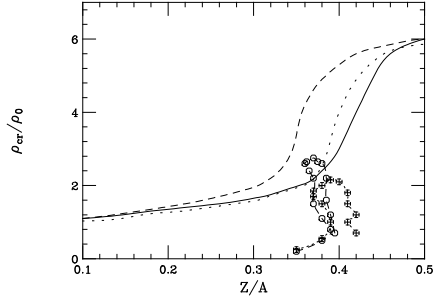}
\vskip -0.3cm
\caption{
Variation of the transition density with proton fraction for various
hadronic $EoS$ parameterizations. Dotted line: $GM3$ $RMF$-model
\cite{GlendenningPRL18};
 dashed line: $NL\rho$ ; solid line: $NL\rho\delta$ . 
For the quark $EoS$: $MIT$ bag model with
$B^{1/4}$=150 $MeV$.
The points represent the path followed
in the interaction zone during a semi-central $^{132}$Sn+$^{132}$Sn
collision at $1~AGeV$ (circles) and at $300~AMeV$ (crosses). 
}
\vskip -1.0cm
\label{rhodelta}
\end{center}
\end{figure}

\begin{figure}[t] 
\centering
\includegraphics[angle=-90,width=8.0cm]{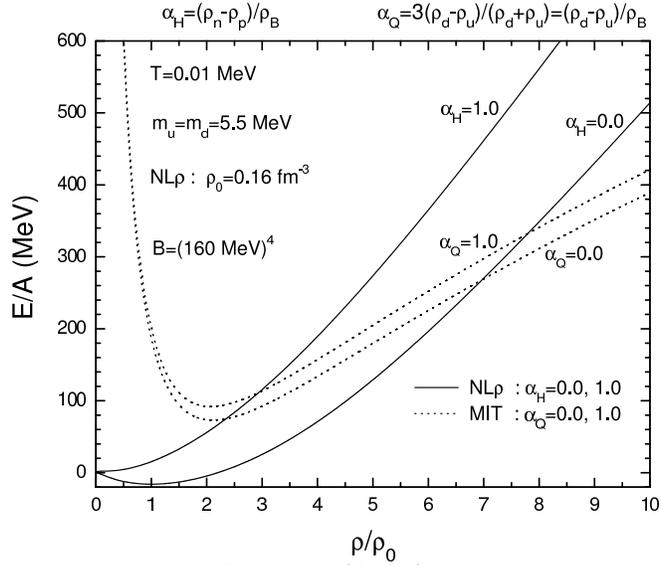} 
\vskip 1.0cm
\caption{$EoS$ of Simmetric/Neutron Matter: Hadron ($NL \rho$), solid lines,
vs. Quark (MIT-Bag), dashed lines. $\alpha_{H,Q}$ represent the isospin 
asymmetry parameters respectively of the hadron,quark matter:
$\alpha_{H,Q}=0$, Symmetric Matter; $\alpha_{H,Q}=1$, Neutron Matter.
}
\label{isoparton} 
\end{figure} 

In this way we get the $binodal$ surface which gives the phase coexistence 
region
in the $(T,\rho_B,\rho_3)$ space.
For a fixed value of the
conserved charge $\rho_3$ 
 we will study the boundaries of the mixed phase
region in the $(T,\rho_B)$ plane. 
In the hadronic phase the charge chemical potential is given by
$
\mu_3 = 2 E_{sym}(\rho_B) \frac{\rho_3}{\rho_B}\, .
$ 
Thus, we expect critical densities
rather sensitive to the isovector channel in the hadronic $EoS$.

In Fig.~\ref{rhodelta}  we show the crossing
density $\rho_{cr}$ separating nuclear matter from the quark-nucleon
mixed phase, as a function of the proton fraction $Z/A$.  
We can see the effect of the
$\delta$-coupling towards an $earlier$ crossing due to the larger
symmetry repulsion at high baryon densities.
In the same figure we report the paths in the $(\rho,Z/A)$
plane followed in the c.m. region during the collision of the n-rich
 $^{132}$Sn+$^{132}$Sn system, at different energies. At
$300~AMeV$ we are just reaching the border of the mixed phase, and we are
well inside it at $1~AGeV$. 

We can expect a {\it neutron trapping}
effect, supported by statistical fluctuations as well as by a 
symmetry energy difference in the
two phases.
In fact while in the hadron phase we have a large neutron
potential repulsion (in particular in the $NL\rho\delta$ case), in the
quark phase we only have the much smaller kinetic contribution.
Observables related to such neutron ``trapping'' could be an
inversion in the trend of the formation of neutron rich fragments
and/or of the $\pi^-/\pi^+$, $K^0/K^+$ yield ratios for reaction
products coming from high density regions, i.e. with large transverse
momenta.  

\begin{figure}
\centering
\includegraphics[angle=-90,width=8.0cm]{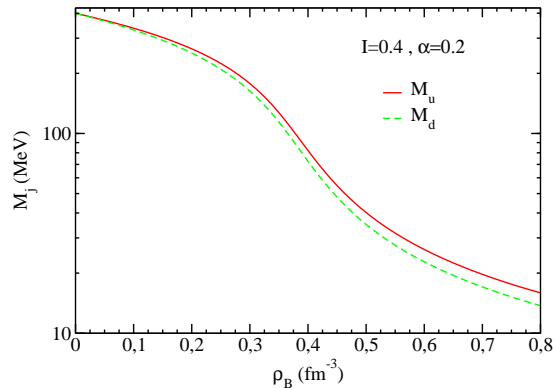} 
\caption{$NJL$ results with flavor mixing $\alpha=0.2$ and isospin 
asymmetry $(N-Z)/A=0.4$: costituent u- d- masses vs.baryon density.
}
\vskip -0.5cm
\label{MuMd} 
\end{figure} 

\vskip 0.5cm
\noindent
{\it Isospin in Effective Partonic Models}

From the above discussion it appears extremely important to include the 
Isospin degree of freedom in any effective approach to the QCD dynamics. 
This can be easily performed in a two-flavor $NJL$ model \cite{NJL} 
where the isospin asymmetry can be included in a flavor-mixing picture 
\cite{frank03} via a Gap Equation like 
$M_i=m_i-4G_1 \Phi_i-4G_2 \Phi_j$, $i \not= j,(u,d)$ , where the 
$\Phi_{u,d}=<\bar u u>,<\bar d d>$ are the two (negative) condensates 
and $m_{u,d}=m$
the (equal) current masses. Introducing explicitily a flavor mixing, i.e.
the dependence of the constituent mass of a given flavor to both condensate,
via $G_1=(1-\alpha) G_0, G_2= \alpha G_0$ we have the coupled equations
\begin{eqnarray}
M_u=m - 4 G_0 \Phi_u - 4 \alpha G_0 (\Phi_u - \Phi_d), \nonumber \\ 
M_d=m - 4G_0 \Phi_u + 4 (1-\alpha) G_0 (\Phi_u - \Phi_d).
\label{mix}
\end{eqnarray}
For $\alpha=1/2$ we have back the usual NJL ($M_u=M_d$), while small/large
mixing is for $\alpha \Rightarrow 0$/$\alpha \Rightarrow 1$ respectively.

In neutron rich matter $\mid \Phi_d \mid$ decreases more rapidly due to the 
larger $\rho_d$ and so $(\Phi_u -\Phi_d)<0$. In the ``realistic'' small mixing 
case, see also \cite{frank03,shao06}, we will get a definite $M_u>M_d$ 
splitting at high baryon density (before the chiral restoration). 
This expectation is nicely confirmed by a full calculation \cite{plum_tesi}
of the coupled gap equations with standard parameters (same as in 
ref.\cite{frank03}). The results are shown in Fig.\ref{MuMd}.

All that represents a more fundamental confirmation of the $m^*_p>m^*_n$ 
choice in the hadron phase,
as suggested by the effective $QHD$ model with the isovector scalar $\delta$ 
coupling, see before and \cite{liubo02}. 
However this can represent just a very first step towards a more complete
treatment of isovector interactions in effective partonic models, of large 
interest for the discussion of the phase transition at high densities.
We can easily see that the mass splitting effect is not changing much the 
symmetry energy in the quark phase addressed before. However confinement 
is still 
missing in these mean field models. Stimulating new perspectives are open.

\vskip -1.0cm
\section{Perspectives}
We have shown that {\it violent} collisions of n-rich heavy ions 
from low to relativistic energies
can bring new information on the isovector part of the in-medium interaction, 
qualitatively different from equilibrium
$EoS$ properties. We have presented quantitative results 
in a wide range of beam energies.
At low energies we see isospin effects on the dissipation in fusion and deep 
inelastic collisions, at Fermi and Intermediate energies the 
Iso-EoS sensitivity of the isospin dynamics in fragment reactions and in 
collective flows.
 
We have shown that meson production in n-rich heavy ions collisions
at intermediate energies
can bring new information on the isovector part of the in-medium interaction
at high baryon densities.
Important non-equilibrium effects for particle production are stressed.
Finally the possibility of observing
precursor signals of the phase transition to a mixed hadron-quark matter
at high baryon density is suggested.

10 years after the first studies of Iso-EoS effects on reactions 
\cite{colonnaPRC57} the symmetry energy case is still open. 
The results presented in this "Isospin Journey" appear 
very promising for 
the possibility of exciting new data from dissipative collisions
with radioactive beams.

\vskip 1.0cm
\noindent
{\bf Aknowledgements}

We warmly thank A.Drago and A.Lavagno for the  
collaboration on the
mixed hadron-quark phase transition at high baryon and isospin density.

One of authors, V. B. thanks for warm hospitality at Laboratori
Nazionali del Sud, INFN. This work was supported in part by the Romanian
Ministery for Education and Research under the contracts PNII, No.
ID-946/2007.

                                    %



\end{document}